\documentstyle[12pt,psfig]{article}

\begin{document}
\title{A possible $0^{-+}$ glueball candidate X(1835)}
\author{Bing An Li\\
Department of Physics and Astronomy, University of Kentucky\\
Lexington, KY 40506, USA\\}

\maketitle

\begin{abstract}
The possibility of X(1835) discovered by BESII as a $0^{-+}$ glueball is studied in this paper.
The decay rates of $X(1835)\rightarrow p\bar{p}, VV$ are associated with 
gluon spin contents of 
proton and vector mesons. Estimations of the decay rates of 
$X(1835)\rightarrow VV, \gamma V, \gamma\gamma$
and the cross sections of $\gamma\gamma\rightarrow X(1835)\rightarrow f$ and $h_1+h_2
\rightarrow X(1835)+...$ are presented.
\end{abstract}

\newpage
The BESII have reported a discovery of a new resonance X(1835) in
$J/\psi\rightarrow\gamma X, X\rightarrow\eta'\pi^+\pi^-$[1].
\[M=1833.7\pm6.1\pm2.7 MeV,\;\;\Gamma=67.7\pm20.3\pm7.7 MeV,\]
\[BR(J/\psi\rightarrow\gamma X)BR(X\rightarrow\eta'\pi^+\pi^-)=(2.2\pm0.4\pm0.4)
10^{-4} \]
are determined.
A narrow enhancement near $2m_p$ in
the invariant mass spectrum of $p\bar{p}$  
of $J/\psi\rightarrow\gamma p\bar{p}$ decays has been reported by BES[2].
No similar structure is seen in $J/\psi\rightarrow\pi^0 p\bar{p}$.
In the range of $M_{p\bar{p}}\leq 1.9GeV$ the angular distribution is consistent with
production of a pseudoscalar or scalar meson in $J/\psi$ radiative decays.
If it is interpreted as a $0^{-+}$ resonance the mass and the width are determined to be  
\[M=1859^{+3}_{-10}(stat)^{+5}_{-25}(syst)MeV\;\;\;\Gamma<30MeV\]
at $90\%$ level[2]. 
\[BR(J/\psi\rightarrow\gamma X)BR(X\rightarrow p\bar{p})=(7.0\pm0.4^{+1.9}_{-0.8})10^{-5}.\]
is determined.
In recent three talks[3] of BES for a $0^{-+}$ X the estimation 
\(BR(J/\psi\rightarrow\gamma X)\sim(0.5-2)\times10^{-3},\;\;\;
BR(X\rightarrow p\bar{p})\sim(4-14)\%\)
are presented.
BR($X\rightarrow p\bar{p}$) is very large.

The masses of the two structures observed in both $\gamma\eta'\pi^+\pi^-$ and $\gamma p\bar{p}$
channels are overlap and $0^{-+}$ quantum number for the resonance in $\eta'\pi^+\pi^-$
channel is possible. A question arises if they are the same state. 
In Ref.[1] an argument is presented 
if the final state 
interaction is included in the fit of the $p\bar{p}$ mass spectrum, the width of 
the resonance observed in $\gamma p\bar{p}$ channel will become larger. Therefore, 
the X observed in both $\gamma p\bar{p}$ 
and 
$\gamma\eta'\pi^+\pi^-$ channels could be the same state and it is named as X(1835)
in Ref.[1]. Various possiblities of the nature of X(1835), such as $p\bar{p}$ bound
state, glueball, and others, are investigated[4].

In this paper the possibility of X(1835) as a candidate 
of $0^{-+}$ glueball[4] is further investigated.  
In Ref.[5] it has been pointed out that $J/\psi\rightarrow\gamma+gg, gg\rightarrow hadrons$ 
provide an 
important search ground for glueballs. To emphasize the importance of 
$J/\psi\rightarrow\gamma+gg$ in search for glueballs it is necessary to mention that
there is another kind of $J/\psi$ radiative decays. 
According to the VMD[6], there are substitutions between photon and vector mesons
\begin{equation}
\rho\rightarrow{1\over2}egA,\;\;\omega\rightarrow{1\over6}egA,\;\;\;
\phi\rightarrow-{\sqrt{2}\over6}egA,
\end{equation}
where \(g=0.39\) is determined by fitting $\rho\rightarrow ee^+$[6].
Because of the VMD 
another kind of radiative decays
$J/\psi\rightarrow ggg, ggg\rightarrow\gamma+hadrons$
are obtained from $J/\psi\rightarrow ggg, ggg\rightarrow \rho(\omega,\phi)+hadrons$,
which are named as VMD-like radiative decays and $J/\psi\rightarrow\gamma+gg, gg\rightarrow 
hadrons$
are named as QCD-like radiative decays. 
Comparing with the QCD-like radiative decays, the VMD-like radiative decays are 
suppressed by 
$O(\alpha^2_s)$. 
Glueballs are more likely produced in 
QCD-like radiative decays with larger branching ratio and less likely in both VMD-like 
radiative decays and $J/\psi\rightarrow V+hadrons$.

$J/\psi\rightarrow\gamma\pi^0$ is an example of the VMD-like radiative decays.
In Ref.[7] BR($J/\psi\rightarrow\gamma\pi^0$) is via the VMD(1) calculated from
the effective Lagrangian of $J/\psi\rightarrow\rho\pi$. BR($J/\psi\rightarrow\gamma\pi^0$)
is small and theory agrees with data well. 
$J/\psi\rightarrow\gamma\sigma(f_0(600))$ is the second example.  
In Ref.[7], using an effective Lagrangian of $J/\psi\rightarrow\omega\sigma$ and 
the VMD(1), very small decay rate of $J/\psi\rightarrow\gamma\sigma$ is
obtained. 
$\sigma$($f_0$(600)) has not been found in $J/\psi$ radiative decays[8]. 
Theory agrees with the data. $J/\psi\rightarrow\gamma\sigma$ is
VMD-like radiative decay. Both $\pi^0$ and $\sigma$ mesons are made of quarks. 

The U(1) anomaly of $\eta'$[9] is known for a long time and $\eta'$ meson is strongly coupled 
to gluons.
The quark components of $\eta'$ meson contribute only about ${1\over3}$ of $m^2_{\eta'}$.
It is known that the quark components dominant the structure of $\eta$ meson. The data[10]
show that $BR(J/\psi\rightarrow\omega(\phi)\eta)>
BR(J/\psi\rightarrow\omega(\phi)\eta')$ and $BR(J/\psi\rightarrow\gamma\eta)<
BR(J/\psi\rightarrow\gamma\eta')$. 
Obviously, the VMD-like of $J/\psi$ radiative decays is not
responsible for large $BR(J/\psi\rightarrow\gamma\eta')$.
These facts show that $\eta'$ contains 
considerable gluon components. Assuming the gluon components of $\eta'$ is responsible
for QCD-like $J/\psi\rightarrow\gamma\eta'$ decay, 
following ratio is obtained in Ref.[11]
\(\frac{BR(J/\psi\rightarrow\gamma\eta')}{BR(J/\psi\rightarrow\gamma\eta)}=5.1\)
which is consistent with data. In Ref.[12] two gluon components in $\eta$ and 
$\eta'$ are estimated.

The BR($J/\psi\rightarrow\rho(\omega,\phi)p\bar{p}$)
are measured to be $<3.1\times10^{-4}, (1.30\pm0.25)\times10^{-3}, (4.5\pm1.5)
\times10^{-5}$ respectively[10]. If $J/\psi\rightarrow\gamma
p\bar{p}$ is VMD-like radiative decay,
using the VMD(1), $BR(J/\psi\rightarrow\gamma 
p\bar{p})$ can be estimated from BR($J/\psi\rightarrow Vp\bar{p}$).   
BR($J/\psi\rightarrow\gamma p\bar{p}$)=$7\times10^{-5}$[2] 
cannot be obtained.
Therefore, larger BR($J/\psi\rightarrow\gamma p\bar{p}$) indicates $J/\psi\rightarrow\gamma X(1835),
X(1835)\rightarrow p\bar{p}$ is QCD-like radiative decay and 
like $\eta'$ X(1835) contains substantial gluon
components.

There is important gluon components in $\eta'$
meson and the $2.2\times10^{-4}$ branching ratio makes 
$J/\psi\rightarrow\gamma X(1835),
X(1835)\rightarrow\eta'\pi^+\pi^-$ a QCD-like radiative decay. 
On the other hand,
the branching ratios of $J/\psi\rightarrow\omega(\phi)+X(1835), 
X(1835)\rightarrow
\eta'\pi\pi$ should be much smaller.

The two decay modes of X(1835)($p\bar{p}$ and $\eta'\pi^+\pi^-$) indicate that X(1835) is
strongly coupled to two gluons. A possible explanation is that X(1835) is a $0^{-+}$ 
glueball.
The study on glueballs has a long history. In Ref.[13] the spectrum of light glueballs 
and pseudoscalar glueball have been studied. The studies of scalar glueballs are 
presented in Ref.[14]. Lattice gauge calculation[15] predicts the spectrum of glueballs.
In this paper effective Lagrangians are constructed and used to study X(1835)  
\begin{eqnarray}
\lefteqn{{\cal L}_{JX\gamma}=eg_{JX\gamma}{1\over m_J}X\epsilon_{\mu\nu\alpha\beta}
\partial^\mu J^\nu\partial^\alpha A^\beta,}\\
&&{\cal L}_{Xg}=\frac{g_{Xg}}{m_X}X\epsilon^{\mu\nu\alpha\beta}F^i_{\mu\nu}
F^i_{\alpha\beta},
\end{eqnarray}
where $F^i_{\mu\nu}$ is the strength tensor of gluon fields. In $J/\psi\rightarrow\gamma 
X(1835)$ X(1835) couples to gluons. Eq.(3) is the effective Lagrangian of the coupling
between X(1835) and gluons.
Using Eq.(2), it is obtained
\begin{equation}
\Gamma(J/\psi\rightarrow\gamma X)=\frac{\alpha}{24}g^2_{JX\gamma}m_J
(1-{q^2\over m^2_J})^3,
\end{equation}
where q is the momentum of X(1835). Based on the value of $BR(J/\psi\rightarrow\gamma X)$[3],
\(g^2_{JX\gamma}=(0.18-0.74)\times10^{-3}\) is estimated.
A glueball can couple to the gluons of a hadron directly. 
The effective Lagrangian ${\cal L}_{Xg}$ is essential 
in determining the glueball nature of X(1835). 
Using Eq.(3),
it is obtained
\begin{equation}
<p\bar{p}|s|X>=i(2\pi)^4\delta(p-k_1-k_2){1\over\sqrt{2m_X}}g_{Xg}{1\over m_X}
<p\bar{p}|F\tilde{F}|0>,
\end{equation}
where $F\tilde{F}=\epsilon^{\mu\nu\alpha\beta}F^i_{\mu\nu}
F^i_{\alpha\beta}$. It is known that this quantity is related to gluon spin 
operator $a_\mu$[16]
\begin{eqnarray}
\lefteqn{F\tilde{F}=-2\partial_\mu a^\mu,\;\;\;
a_\mu=-\epsilon^{\mu\nu\alpha\beta}\{F^i_{\nu\alpha}A^i_\beta-{g\over 3}C_{ijk}
A^i_{\nu}A^j_{\alpha}A^k_{\beta}\},}\\
&&<p\bar{p}|F\tilde{F}|0>=-2i(k_1+k_2)^\mu<p\bar{p}|a_\mu|0>,
\end{eqnarray}
where $A^i_\mu$ is gluon field.
The matrix element of gluon spin content of a proton is expressed as[16]
\begin{equation}
<p|a_\mu|p>=\Delta G\bar{u}\gamma_\mu\gamma_5 u,
\end{equation}
where $\Delta G$ is the gluon spin content of proton. In the region of 
X(1835)-resonance the momenta of proton and antiproton produced in the decay 
of X(1835) are very low. Using cross symmetry and Eq.(6), it is determined
\begin{equation}
<p\bar{p}|F\tilde{F}|0>=-4i\Delta G m_N\bar{u}\gamma_5 v.
\end{equation}
The diagonal matrix elements of $a_\mu$ are gauge invariant. For nondiagonal matrix elements
of $a_\mu$ there are other terms. However, the divergence of these terms is zero and they don't
affect Eq.(9).
$\Delta G$ is an important quantity in understanding the proton spin. There are very
active experimental projects on measuring it. In Ref.[17] following data are quoted
\begin{equation}
\Delta G(1GeV^2)=0.99^{+1.17 +0.41 +1.43}_{-0.31 -0.22 -0.45}(SMC),\;\;\;
\Delta G(5GeV^2)=1.6\pm1.1(E155).
\end{equation}
\begin{equation}
\Gamma(X\rightarrow p\bar{p})={2\over\pi\sqrt{q^2}}g^2_{Xg}m^2_N(\Delta G)^2
(1-{4m^2_N\over q^2})^{{1\over2}}
\end{equation}
is derived, where q is the momentum of X(1835).
Eqs.(4,11) lead to  
\begin{equation}
\frac{d\Gamma}{dq^2}(J/\psi\rightarrow\gamma X, X
\rightarrow p\bar{p})=\frac{\sqrt{q^2}}{\pi}\frac{\Gamma(J/\psi\rightarrow\gamma X)
\Gamma(X\rightarrow p\bar{p})}{(q^2-m^2_X)^2+q^2\Gamma^2_X},
\end{equation}
where $q=k_1+k_2$, $k_1$ and $k_2$ are momenta of p and $\bar{p}$ respectively.
BES have measured the distribution of the decay rates of $J/\psi\rightarrow\gamma p\bar{p}$
[2]. 
Choosing \(\Gamma_X=50MeV\), Eq.(12) fits the data[2] well(Fig.1) and the value 
of $\Gamma_X$  
is consistent with the one determined from $J/\psi\rightarrow\gamma\eta'\pi\pi$. 
\begin{figure}
\begin{center}
\psfig{figure=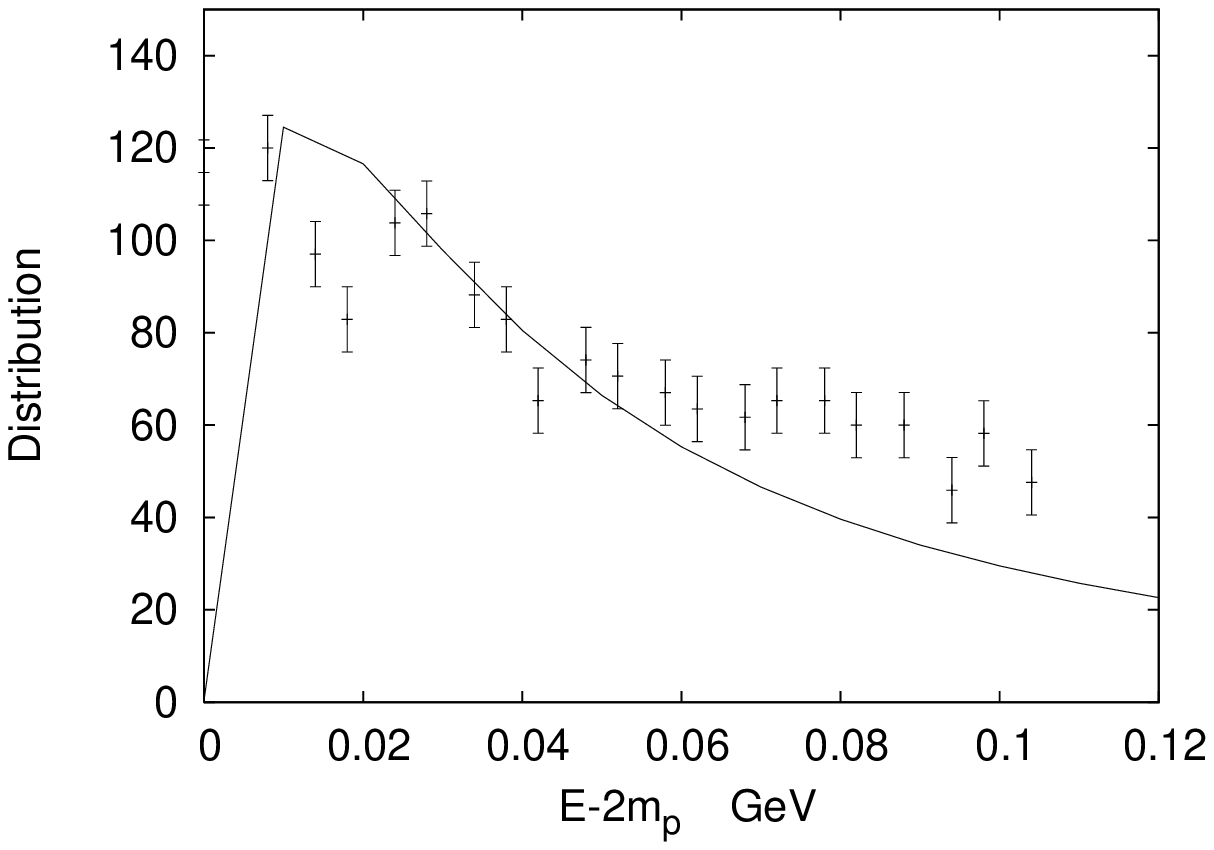}
FIG. 1.
\end{center}
\end{figure}

The glueball explanation of X(1835) supports that the two structures found in both
$\gamma p\bar{p}$ and $\gamma\eta'\pi\pi$ originate is one resonance state.
From Fig.1 it can be seen that outside of the resonance theory decreases faster than data.
Besides the resonance there are other processes which contribute to $J/\psi\rightarrow\gamma 
p\bar{p}$. Study of these processes are beyond the scope of this paper.
Eq.(9) shows that the glueball X(1835) is coupled to the gluons of proton directly and 
there is no
QCD suppression. Therefore, large BR($X\rightarrow p\bar{p}$) is understood
qualitatively. Using BR($X\rightarrow p\bar{p}$)[3], it is estimated
\(g_{Xg}\Delta G=0.27\sim 0.51\). If taking \(\Delta G\sim 1\), \(g_{Xg}=0.27\sim 0.51\).
The QCD-like radiative decays predict 
\(\Gamma(J/\psi\rightarrow\gamma n\bar{n})=\Gamma(J/\psi\rightarrow\gamma p\bar{p})\).

The decay $J/\psi\rightarrow\gamma X(1835)$
is similar to $J/\psi\rightarrow\gamma\eta'$ in which only the gluon components of $\eta'$ 
contribute[11].   
Approximately 
\begin{equation}
\frac{\Gamma(J/\psi\rightarrow\gamma X(1835))}{\Gamma(J/\psi\rightarrow\gamma\eta')}
\sim{1\over sin^2\theta}
({k_X\over k_{\eta'}})^3,
\end{equation}
where $sin\theta$ is defined in \(|eta'>=cos\theta|q\bar{q}>+sin\theta|gg>\), 
\(k_X={m_J\over2}(1-{m^2_X\over m^2_J})\), and \(k_{\eta'}={m_J\over2}(1-{m^2_{\eta'}
\over m^2_J})\).
Large BR($J/\psi\rightarrow\gamma\eta'$) indicates that $sin\theta$
is not small. Both BR($J/\psi\rightarrow\gamma X(1835))$
and BR($J/\psi\rightarrow\gamma\eta')$ are at the same order of magnitude($\sim10^{-3}$) 
which is
consistent with the estimation made in Ref.[3]. The gluon components of $\eta$ are very small, 
therefore,
glueball X(1835) predicts that
$BR(X(1835)\rightarrow\eta'\pi\pi)$ is much greater than BR($X(1835)\rightarrow\eta\pi\pi$).

The hadronic decays of $J/\psi$ are the processes $J/\psi\rightarrow ggg, ggg\rightarrow 
quark\; pairs\rightarrow hadrons$.
The small branching ratios of $J/\psi\rightarrow\omega(\phi)\eta'$ are understood qualitatively
by the fact that $\eta'$ contains important gluon components. Similarly, 
very small BR($J/\psi\rightarrow\omega(\phi)X(1835)$)
is a prediction of the glueball picture of X(1835).
The isospin of two pions of 
$J/\psi\rightarrow\gamma X(1835), X(1835)\rightarrow\eta'\pi^+\pi^-$ is zero. 
Naturally, \(BR(\pi^0\pi^0)={1\over2}BR(\pi^+\pi^-)\) 
should be expected. 

Besides $\eta$ and $\eta'$ there are other three $0^{-+}$ states: $\eta(1295), \eta(1405)$,
and $\eta(1475)$[10]. Especially $BR(J/\psi\rightarrow\gamma\eta(1405/1475))$
is compatible with $J/\psi\rightarrow\gamma\eta'$ and $J/\psi\rightarrow\phi\eta(1405)$ 
is smaller[10]. 
In Ref.[29] it is argued that there is a $0^{-+}$ glueball candidate among them.
It is interesting to measure
BR$(X(1835)\rightarrow\eta(1295/1405/1475)+\pi\pi$) to see whether a larger BR is among 
them. 
The measurements
BR$(J/\psi\rightarrow\omega(\phi)\eta(1275/1405/1475))$ are interesting too.  

$J/\psi\rightarrow\gamma X(1835), X(1835)\rightarrow\rho\rho, \omega\omega, K^*\bar{K}^*$ 
are other possible decay channels. 
A broad $\rho^0\rho^0$ enhancement at $1650\pm200$MeV 
in $J/\psi\rightarrow\gamma\rho\rho$ with 
$\Gamma=200\pm100$ MeV has been reported[18]. In Ref.[19] it has been indicated that about 
$50\%$ 
of the decays is due to a $0^-$ resonance and
\(BR(J/\psi\rightarrow\gamma X(1.5-1.9 GeV, 0^-))BR(X\rightarrow\rho^0\rho^0)=(7.7\pm3.0)
10^{-4}\).
The $0^{-+}$ component appears strongly in BES data of $J/\psi\rightarrow\gamma+2(\pi^+\pi^-)$[20].
The $\omega\omega$ events reveal the similar structure. In Ref.[21] MARK III has reported that 
a dominant $0^{-+}$ component
accounts for $55\%$ of the data of $J/\psi\rightarrow\gamma K^*\bar{K}^*$. In this reaction    
BES[22] has found a broad $0^{-+}$ $K^*\bar{K}^*$ resonance with $M=1800\pm100$MeV and 
$\Gamma=500\pm200$MeV.
In Ref.[23] the MARK III data of $J/\psi\rightarrow\gamma \eta\pi\pi, \rho\rho, \omega\omega, 
\phi\phi$, and $K^*\bar{K}^*$ have been analyzed and a very broad $0^-$ component 
in $K^*\bar{K}^*$ channel with mass of
1750-2190 MeV and a width of 1 GeV is found. 
In Ref.[24] the production of $2^{++}$ $q^2\bar{q}^2$
states in $J/\psi\rightarrow\gamma+VV$ has been studied. The resonance structures observed
in $J/\psi\rightarrow\gamma+VV$ are more complicated. It is possible that $0^{-+}$ glueball X(1835)
is produced in $J/\psi\rightarrow\gamma+X(1835), X(1835)\rightarrow VV$. 

The decays rates of $X(1835)\rightarrow VV$ can be calculated by using Lagrangian(3)
\begin{eqnarray}
\lefteqn{<VV|s|X>=i(2\pi)^4\delta^4(p-k1-k2){1\over\sqrt{2m_X}}{g_{Xg}\over m_X}
<VV|F\tilde{F}|0>,}\nonumber \\
&&<VV|F\tilde{F}|0>=2ip^\mu<VV|a_\mu|0>,
\end{eqnarray}
where p, $k_1$ and $k_2$ are momentum of X and V's respectively.  
It is known that gluon spin content is important for proton spin.
It is necessary to investigate whether gluon spin plays any role in understanding
the spin of a vector meson.
We use a chiral model of 
pseudoscalar, vector, and axial-vector mesons[6], which is very successful phenomenologically,
to calculate the quark spin content of a vector meson, (for example, $\rho$ meson),
\[<\rho^i|\bar{\psi}\gamma_\mu\gamma_5\psi|\rho^i>=(\Delta u+\Delta d)<\rho^i|s_\mu|\rho^i>,\]
where $s_\mu$ is the spin operator of vector meson(see below). 
The $\rho$ meson vertex is defined as[6]
\({1\over g}\bar{\psi}\tau^i\gamma_\mu\psi\rho^i\),
where g has been shown in Eq.(1) and the
value of g determines $\Gamma_\rho=150MeV$. In the chiral limit g and $f_\pi$
are the two parameters in this chiral theory. 
The calculation shows 
\({1\over2}(\Delta u+\Delta d)={1\over 2\pi^2 g^2}=0.34\).
$\Delta s=0$ is obtained in the leading order in $N_C$ expansion. The quark spin content of 
$\rho$ meson is only ${1\over 3}$ the total spin of $\rho$ meson. 
\[<\rho^i \rho^i|\bar{\psi}\gamma_\mu\gamma_5\psi|0>=(\Delta u+\Delta d)<\rho^i 
\rho^i|s_\mu|0>\]
is obtained too and
cross symmetry is satisfied.
It is reasonable that gluon spin
content is important part of the spin of a vector meson.
After using the cross symmetry, the gluon spin content is defined as
\begin{equation}
<VV|a_\mu|0>=\Delta G_V<VV|s^V_\mu|0>,
\end{equation}
where $\Delta G_V$ is the gluon spin content of vector meson and
$s^V_\mu$ is the spin operator of vector meson
\begin{equation}
s^V_\mu=-\epsilon^{\mu\nu\alpha\beta}\{V^a_{\nu\alpha}V^a_\beta-{g\over3}
C_{abc}V^a_\nu V^b_\alpha
V^c_\beta\},\;\;\;
V^a_{\mu\nu}=\partial_\mu V^a_\nu-\partial_\nu V^a_\mu+gc_{abc}V^b_\mu V^c_\nu,
\end{equation}
where a is the index of flavor SU(3), $V^a_{\mu\nu}$
is the strength tensor of vector field $V^a_\mu$.
The measurements of the decay rate of $X(1835)\rightarrow VV$ will provide important 
information 
of gluon spin content of vector meson. Of course if the gluon spin content is very small the quark
operators produced by $F\tilde{F}$ must be taken into account and  
a suppression at $O(\alpha^2_s)$ for the decay rate of $X(1835)\rightarrow VV$
is the consequence and BR($X(1835)\rightarrow VV$) should be smaller. 
It is derived
\begin{equation}
\Gamma(X\rightarrow VV)={2f_V\over \pi}g^2_{Xg}(\Delta G_V)^2\sqrt{q^2}
(1-{4m^2_V\over q^2})^{{3\over2}},
\end{equation}
where q is the momentum of X, \(f_V=1,{1\over2},{1\over2},1,1\) for \(V=\rho^{\pm}, \rho^0, 
\omega, K^{*\pm}, 
K^{*0}\) 
respectively. The distribution of the decay $J/\psi\rightarrow\gamma X, X\rightarrow VV$ is
obtained
\begin{equation}
\frac{d\Gamma}{dq^2}(J/\psi\rightarrow\gamma X,X\rightarrow VV)=\frac{\sqrt{q^2}}{\pi}
\frac{\Gamma(J/\psi\rightarrow\gamma X)
\Gamma(X\rightarrow VV)}{(q^2-m^2_X)^2+q^2\Gamma^2_X}.
\end{equation}
In the region of the resonance of X(1835) 
we obtain
\begin{eqnarray}
\lefteqn{BR(J/\psi\rightarrow\gamma+X, X\rightarrow\rho\rho)=1.05\times(0.13-1.89)10^{-3}
\Delta G^2_V,}\nonumber \\
&&BR(J/\psi\rightarrow\gamma+X, X\rightarrow K^*\bar{K}^*)=1.2\times(0.13-1.89)10^{-4}
\Delta G^2_V.
\end{eqnarray}
The brancing ratio of $\omega\omega$ decay mode is ${1\over3}$ of $\rho\rho$. 
If taking $\Delta G_V\sim 0.6$, the results are compatible with data[10] of 
BR($J/\psi\rightarrow\gamma
VV$).
Using the VMD, the decay rates of $J/\psi\rightarrow\gamma+X, X\rightarrow\gamma+V$ can be 
calculated
\begin{eqnarray}
\lefteqn{\frac{d\Gamma(J/\psi\rightarrow\gamma+X, X\rightarrow\gamma\rho)}{dq^2}=
\frac{\sqrt{q^2}}{\pi}\frac{\Gamma(J/\psi\rightarrow\gamma X)
\Gamma(X\rightarrow\gamma \rho)}{(q^2-m^2_X)^2+q^2\Gamma^2_X},}\\
&&\Gamma(X\rightarrow\gamma \rho)=2\alpha g^2\sqrt{q^2}(1-\frac{m^2_\rho}{q^2})^3g^2_{Xg}
\Delta G^2_V,\;\;
\Gamma(X\rightarrow\gamma\omega)={2\over9}\Gamma(X\rightarrow\gamma\rho),\\
&&\Gamma(X\rightarrow\gamma\phi)={4\over9}\Gamma(X\rightarrow\gamma\rho)\frac
{(1-\frac{m^2_\phi}{q^2})^3}{(1-\frac{m^2_\rho}{q^2})^3},
\end{eqnarray}
Numerical results are
\begin{eqnarray}
\lefteqn{BR(J/\psi\rightarrow\gamma X, X\rightarrow\gamma\rho)=0.88\times(0.13-1.89)10^{-5}
\Delta G^2_V,}\\
&&\frac{BR(J/\psi\rightarrow\gamma X, X\rightarrow\gamma\rho)}
{BR(J/\psi\rightarrow\gamma X, X\rightarrow\rho^0\rho^0)}
=0.83\times10^{-2},\;\;
\frac{BR(J/\psi\rightarrow\gamma X, X\rightarrow\gamma\phi)}
{BR(J/\psi\rightarrow\gamma X, X\rightarrow\gamma\rho^0)}
=0.11\times10^{-2}.
\end{eqnarray}
The measurements of the decay rates of $J/\psi\rightarrow\gamma+X, X\rightarrow K^*\bar{K}^*$ and
$J/\psi\rightarrow\gamma+X, X\rightarrow\gamma\phi$ can test the flavor independence of glueball
X(1835).

The $0^{-+}$ glueball X(1835) can be produced by two photons. 
In Ref.[7] the VMD has been extended to $J/\psi$. Using the substitution
\(J_\mu\rightarrow eg_J A_\mu\),
where $g_J=0.0917$ is determined by fitting the decay rate of $J/\psi\rightarrow ee^+$,
\begin{equation}
{\cal L}^{(1)}_{X\gamma\gamma}=e^2 g_J{g_{JX\gamma}\over m_J}X\epsilon^{\mu\nu\alpha\beta}
\partial_\mu
A_\nu \partial_{\alpha}A_\beta
\end{equation}
is obtained.
The second part of the effective Lagrangian of $X\rightarrow\gamma\gamma$
is obtained from $X\rightarrow VV$ by the substitutions(1)
\begin{equation}
{\cal L}^{(2)}(X\rightarrow\gamma\gamma)={4\over3}e^2 g^2{g_{Xg}\over m_X}\Delta G_V
\epsilon^{\mu\nu\alpha\beta}\partial_\mu A_\nu\partial_\alpha A_\beta.
\end{equation} 
The cross section of $\gamma\gamma\rightarrow X\rightarrow f$ is found
\begin{equation}
\sigma(\gamma\gamma\rightarrow X\rightarrow f)=8\pi\frac{\Gamma(X\rightarrow\gamma\gamma)\Gamma(X\rightarrow f)}
{(q^2-m^2_X)^2+q^2\Gamma^2_X},
\end{equation}
where
\begin{equation}
\Gamma(X\rightarrow\gamma\gamma)=\pi\alpha^2\{g_J g_{XJ\gamma}+{4\over3}{m_J\over m_X}g^2 g_{Xg}
\Delta G_V\}^2\frac{m^3_X}{m^2_J}=1.1(0.31-1.1)keV.
\end{equation}
The contribution of ${\cal L}^{(1)}$ can be ignored.
Because of the gluon spin content of vector meson in $X\rightarrow\gamma\gamma$ there is 
no suppression at $O(\alpha^2_s)$, therefore,
$\Gamma(X\rightarrow\gamma\gamma)$ is not narrow. 
As a matter of fact, $\sigma(\gamma\gamma\rightarrow p\bar{p})$ has been measured 
above 2GeV[25].
X(1835) can be via Primakov effects produced in photoproduction $\gamma p\rightarrow X(1835)p$.
The cross section
\begin{equation}
\frac{d\sigma}{d\Omega}=\Gamma(X\rightarrow\gamma\gamma)\frac{8\alpha Z^2}{m^3_X}
\frac{\beta^3E^4}{Q^4}|F_{em}(Q)|^2 sin^2\theta_m
\end{equation}
can be found in Ref.[26]. The cross section of X(1835) is at the same order of magnitude
as $\pi^0$. On the other hand, X(1835) can via a $\rho$ exchange be produced in 
$\gamma p\rightarrow X(1835)+p$. 
At $q^2\sim m^2_X$ the estimations   
$\sigma(\gamma\gamma\rightarrow X\rightarrow p\bar{p})\sim pb$ and
$\sigma(\gamma\gamma\rightarrow X\rightarrow\eta'\pi^+\pi^-)\sim 
3\sigma(\gamma\gamma\rightarrow X\rightarrow p\bar{p})$ are made.

X(1835) can be produced in the central region of hadron collisions by two gluons. 
In the parton model the cross section of the production
of X(1835) is written as
\begin{equation}
\sigma(h_1+h_2\rightarrow X+...)=\int^1_{x_{1min}}\int^1_{x_{2min}}dx_1 dx_2
\{G^{h_1}_{g_1}(x_1)G^{h_2}_{g_2}(x_2)+G^{h_1}_{g_2}(x_1)G^{h_2}_{g_1}(x_2)\}
\sigma(g_1+g_2\rightarrow X\rightarrow f),
\end{equation}
where $G^h_g(x)$ is the gluon distribution function of hadron and
\(x_{1min}=(m_1+m_2)/S,\;\;\;
x_{2min}=(m_1+m_2)/(x_1 S)\),
$m_1, m_2$ are the masses of the two hadrons, \(S=(p_1+p_2)^2\),
\begin{eqnarray}
\sigma(g_1+g_2\rightarrow f)&=&{\pi\over 2}\frac{\Gamma(X\rightarrow gg)\Gamma(X\rightarrow f)}
{(q^2-m^2_X)^2+q^2\Gamma^2_X},
q^2=x_1 x_2 S,\\
\Gamma(X\rightarrow gg)&=&{4\over \pi}g^2_{Xg}\sqrt{q^2},\\
\frac{d\sigma(h_1 h_2\rightarrow X\rightarrow f)}{dq^2}&=&\int^1_{x_{1min}}{1\over x_1 S} dx_1
\{G^{h_1}_{g_1}(x_1)
G^{h_2}_{g_2}(x_2)+G^{h_1}_{g_2}(x_1)G^{h_2}_{g_1}(x_2)\}\sigma(g_1 g_2\rightarrow X\rightarrow f).
\end{eqnarray}
The gluon distribution functions of nucleon and pion are chosen as
\begin{equation}
G^{\pi^+}_g (x)=2(1-x)^3/x,\;\;\;
G^p_g (x)=2.62(1+3.5x)(1-x)^{5.9}/x.
\end{equation}
Using Eqs.(31-34), the cross sections of the production of X(1835) in hadron collisions are  
estimated. As an example at \(s=750GeV^2\) $\sigma(pp\rightarrow X+..., X\rightarrow p\bar{p})
\sim 35\mu b$, ($BR(X\rightarrow p\bar{p})=0.04$ is taken); $\sigma(pp\rightarrow X+..., 
X\rightarrow \eta'\pi^+\pi^-)\sim 105\mu b$, ($BR(X\rightarrow \eta'\pi^+\pi^-)=0.12$ is taken). 
For $\pi p$ collisions the two corresponding cross sections are 29$\mu b$ and 90 $\mu b$ 
respectively.

X(1835) can be produced in $ee^+\rightarrow J/\psi+X(1835)$[27]. 
In Ref.[28] the cross section of $ee^+\rightarrow J/\psi+glueball(0^{++})$ has been 
calculated to be $\sim fb$
at the energy 10.6GeV. The process $\gamma^*\rightarrow(c\bar{c})(gg)$ involved in this process 
is the same as in $ee^+\rightarrow J/\psi+X(1835)$. It is possible that the cross sections of both processes
could be at the same order of magnitude, $\sigma(ee^+\rightarrow J/\psi+X(1835))\sim fb$ 
at the energy
10.6GeV.

It is necessary to mention the decay $B^+\rightarrow p\bar{p} K^+$ has been 
reported by Belle[30] in 2002.
The mass spectrum of $p\bar{p}$ has a wider distribution. However, as mentioned in Ref.[30]
that $p\bar{p}$ mass spectrum is peaked toward low mass and for $M(p\bar{p})< 2GeV$ 
$BR=(0.9^{+0.42}_{-0.35})10^{-6}$ which is larger than the BR in 2.0-2.2 GeV range. The structure of
the mass spectrum of $p\bar{p}$ of $B^+\rightarrow p\bar{p} K^+$ is more complicated than the one
in $J/\psi\rightarrow\gamma p\bar{p}$[2]. The process $B^+\rightarrow K^+ +c\bar{c},
c\bar{c}\rightarrow gg, gg\rightarrow p\bar{p}$ is a possible contributor of 
$B^+\rightarrow p\bar{p} K^+$. According to the study presented in this paper,
the possibility of the production of X(1835)
in the decay $B^+\rightarrow K^+ X(1835), X(1835)\rightarrow p\bar{p}$ cannot be ruled out.
The measurements $B^+\rightarrow K^+ X(1835), X(1835)\rightarrow\eta'\pi\pi$ are necessary
to establish the production of X(1835) in B decays. Using the data presented
in Refs.[1,2], $BR(B^+\rightarrow K^+ X(1835), X(1835)\rightarrow\eta'\pi^+\pi^-)\sim 3\times
10^{-6}$ is obtained. Based on Eqs.(19) $BR(B^+\rightarrow K^+ X(1835), X(1835)\rightarrow \rho\rho, 
K^*\bar{K}^*)$ are in the range of $10^{-6}$. Detailed studies of these B decays are beyond
the scope of this paper.
 
This work is supported by a DOE grant.

\begin{flushleft}
{\bf Figure Captions}
\end{flushleft}
{\bf FIG. 1.} Distribution of $J/\psi\rightarrow p\bar{p}$ with arbitrary unit.

\end{document}